\documentclass[fleqn,usenatbib]{mnras}

\usepackage{newtxtext,newtxmath}

\usepackage[T1]{fontenc}
\usepackage{ae,aecompl}
\usepackage{longtable}

\usepackage{graphicx}	
\usepackage{amsmath}	
\usepackage[flushleft]{threeparttable}
\usepackage{xcolor}
\newcommand{\teff}{$T_\mathrm{eff}$}
\usepackage{lscape}
\usepackage{ulem}

\title[M dwarfs in VVV b294 field]{M-dwarf stars in the b294 field from the VISTA Variables in the Vía Láctea (VVV)}
\author[P. Cruz et al.]{
Patricia Cruz$^{1,2}$\thanks{E-mail: pcruz@cab.inta-csic.es},
Miriam Cort\'{e}s-Contreras$^{1,2}$,
Enrique Solano$^{1,2}$,
Carlos Rodrigo$^{1,2}$,
\newauthor
Dante Minniti$^{3,4}$,
Javier Alonso-Garc\'{i}a$^{5,6}$,
Roberto K. Saito$^{7}$
\\
$^{1}$Centro de Astrobiolog\'{\i}a (CAB), CSIC-INTA, Camino Bajo del Castillo s/n, E-28692, Villanueva de la Ca\~{n}ada, Madrid, Spain\\
$^{2}$Spanish Virtual Observatory\\
$^{3}$Instituto de Astrof\'{\i}sica, Facultad de Ciencias Exactas, Universidad Andres Bello, Fernández Concha 700, Las Condes,
Santiago, Chile\\
$^{4}$ Vatican Observatory, Vatican City State, V-00120, Italy\\
$^{5}$ Centro de Astronom\'{i}a (CITEVA), Universidad de Antofagasta, Av. Angamos 601, Antofagasta, Chile\\
$^{6}$ Millennium Institute of Astrophysics, Nuncio Monse\~{n}or Sotero Sanz 100, Of. 104, Providencia, Santiago, Chile\\
$^{7}$ Departamento de Física, Universidade Federal de Santa Catarina, Trindade 88040-900, Florianópolis, SC, Brazil\\
}

\date{Accepted 2023 January 30. Received 2023 January 27; in original form 2022 July 14}

\pubyear{2022}

\begin{document}
\label{firstpage}
\pagerange{\pageref{firstpage}--\pageref{lastpage}}
\maketitle

\begin{abstract}
M-dwarf stars are the dominant stellar population in the Milky Way and they are important for a wide variety of astrophysical topics. 
The {\sl Gaia} mission has delivered a superb collection of data, nevertheless, ground-based photometric surveys are still needed to study faint objects. Therefore, the present work aims to identify and characterise M-dwarf stars in the direction of the Galactic bulge using photometric data and with the help of Virtual Observatory tools. 
Using parallax measurements and proper motions from {\sl Gaia} Data Release 3, in addition to different colour-cuts based on VISTA filters, we identify and characterise 7\,925 M-dwarf stars in the b294 field from the Vista Variables in the Vía Láctea (VVV) survey. We performed a spectral energy distribution fitting to obtain the effective temperature for all objects using photometric information available at Virtual Observatory archives. The objects in our sample have temperatures varying from 2800 to 3900 K. We also search for periodic signals in VVV light curves with up to 300 epochs, approximately. 
As a secondary outcome, we obtain periods for 82 M dwarfs by applying two methods: the Lomb-Scargle and Phase Dispersion Minimization methods, independently. These objects, with periods ranging from 0.14 to 34\,d, are good candidates for future ground-based follow up. Our sample has increased significantly the number of known M dwarfs in the direction of the Galactic bulge and within 500\,pc, showing the importance of ground-based photometric surveys in the near-infrared.
\end{abstract}




\section{Introduction}

M dwarfs are the dominant stellar population in the Milky Way \citep{Kroupa01, Chabrier03}. Because of their ubiquity and their lifetimes in the main-sequence as long as the age of the Universe \citep{Laughlin97}, these low-mass, cool dwarfs objects -- with masses from $\sim$0.075 to 0.6 M$_\odot$, depending on the metallicity \citep{Chabrier00}, and with effective temperatures ($T_{\rm eff}$) ranging from 2\,350 to 3\,850\,K, approximately \citep{Pecaut13} -- are important for a wide variety of astrophysical contexts. For instance, they are studied as tracers of the structure, kinematics, and evolution of the Galaxy \citep[e.g.,][]{Scalo86, Bochanski07, Ferguson17} or in the discussion of formation and evolution processes involving stellar objects at the end of the Herzprung-Russell diagram \citep[e.g.,][]{Jeffries04, Stamatellos09, Luhman12}. More recently, M-dwarf stars have been defined as prime targets of exoplanet surveys as possible hosts of Earth-like planets and in the search for life elsewhere in the Galaxy \citep[e.g.,][]{Bonfils13,Reiners18,Wunderlich19}.

M dwarfs emit the bulk of their energy in the near-infrared range of the spectrum. Therefore, large area photometric surveys operating at these wavelengths, such as for instance the Two Micron All Sky Survey (2MASS, \citealt{Skru06}) and the UKIRT Infrared Deep Sky Survey (UKIDSS, \citealt{UKIDSS2007}), have been valuable resources for M dwarf identification and characterisation, in combination with broadband photometry in the visible range. These data typically come from Pan-STARRS DR1 \citep{Kaiser10,Chambers16}, SDSS \citep{York00}, and {\sl Gaia} \citep{Gaia2016} surveys, which have been extensively used for this purpose \citep[e.g.,][]{Cook16,Lodieu17,Bentley19}.

Several works have been devoted to determine the photometric colours observed in the M dwarf regime \citep{West05,Cifuentes2020}. 
All the efforts put on the identification of M dwarfs have generally avoided the Galactic bulge, with plenty of stars, gas and dust. It is in this direction where the public ESO VISTA Variables in the Via Lactea (VVV) survey operates \citep{Minniti10, Minniti18}. Using colour-cuts from VVV b201 field, a field near the bottom-right region of the Galactic bulge, \citet{Rojas-Ayala2014} identified over 23\,300 M-dwarf candidates. These authors also estimated photometric distances, since {\sl Gaia} parallaxes were not available yet, and identified among their sample possible giant contaminants based on a reduced proper motion diagram.  

The {\sl Gaia} mission has delivered a superb collection of data, which include astrometric measurements for almost two billion stars, a smaller set but yet an impressive amount of light curves, low-resolution spectra and radial velocity measurements. Nevertheless, the great majority of them are not available for targets with $G$-band magnitude fainter than $\sim$17\footnote{For details, see {\sl Gaia} documentation available at \url{https://www.cosmos.esa.int/web/gaia/dr3}.}. 
Therefore, ground-based photometric surveys still have an important role in the {\sl Gaia} era, especially for faint objects. The present work aims to identify and characterise M-dwarf stars towards the bulge of the Galaxy using photometric data only and with the help of Virtual Observatory tools.

In this work, we explored the VVV b294 field located in the inner bulge region. The description of the used data is presented in Sect.~\ref{data}. The adopted sample selection methods and the list of M-dwarf candidates are presented in Sect.~\ref{sample}. The stellar characterisation based on the stellar spectral energy distribution is detailed in Sect.~\ref{vosa}. A comparison with M stars from {\sl Gaia} is performed in Sect.~\ref{discuss}. As a secondary outcome, the search for periodicity in the VVV light curves is described in \ref{period}. Finally, our conclusions are presented in Sect.~\ref{concl}.

\section{VVV Data}\label{data}

The VVV Survey \citep[][]{Minniti10,Saito12} is a large ESO public near-IR surveys that mapped the bulge and inner disk of the Milky Way in the near infrared since 2011. The observations were performed using the 4.1m-telescope at the Cerro Paranal Observatory (Chile) and the data reduction was carried
out at the Cambridge Astronomical Survey Unit \citep{Irwin04,Lewis10}. The observations cover the near-infrared $ZYJHK_s$ passbands, up to four magnitudes deeper than 2MASS. In addition, a variability campaign was done from 2010 to 2016, where $K_S$-band light curves were generated with up to 300 epochs \citep[for more details, see][and references therein]{Saito12,Botan21}.

In this work we have focused on the b294 field, which spans among the following coordinates: $3.1^{\circ} < l < 4.7^{\circ}$ and $-3.8^{\circ} < b < -2.5^{\circ}$. 
For our analysis, we use the photometric catalogue for this VVV region by \citet{Alonso18}, extracted using point spread function (PSF)-fitting techniques. 
All the details on image reduction, PSF photometry, and quality flags are described in \citet{Alonso18}. The VVV b294 field cointains a total of $4\,592\,101$ detected point-source objects.

\begin{figure*}
    \includegraphics[width=\textwidth,trim=2cm 1.8cm 2cm 1.5cm, clip]{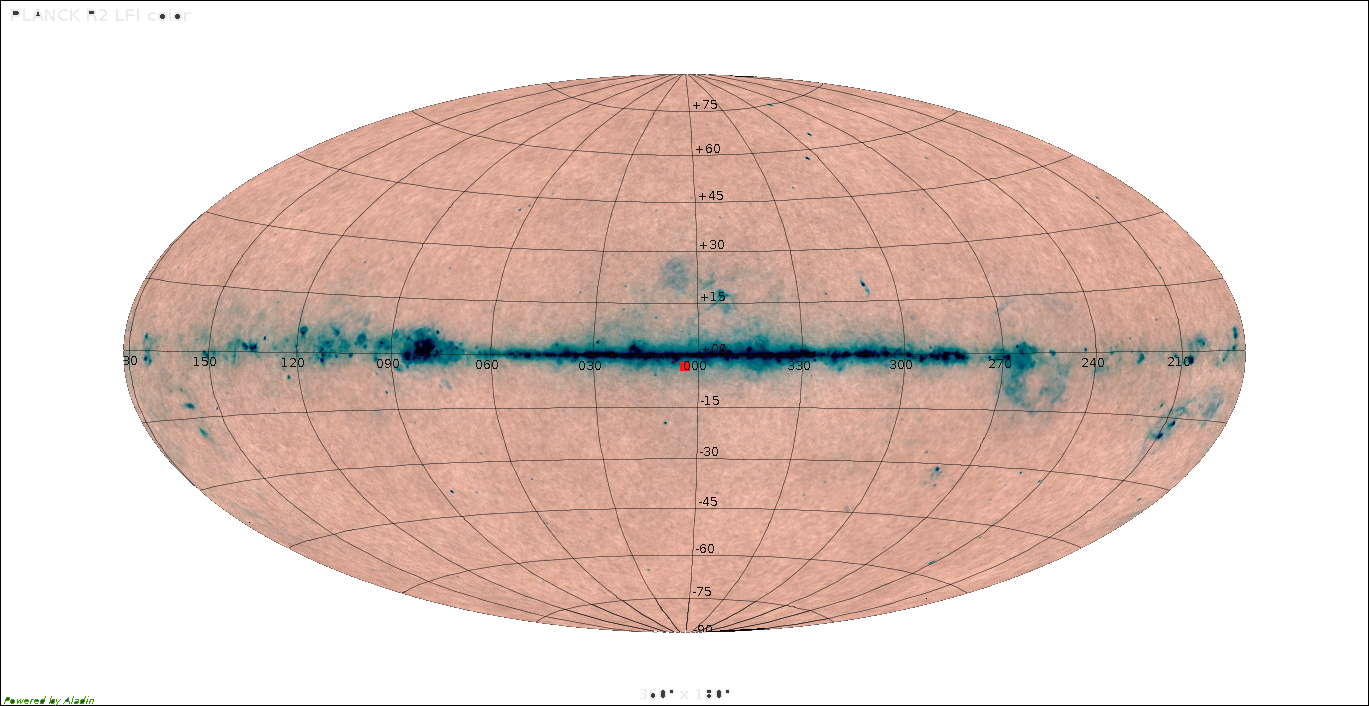}
    \caption{Location of the VVV b294 field (red filled square) in the sky in Galactic coordinates. A Planck map in aitoff projection is displayed in the background.}
    \label{fig_b294skymap}
\end{figure*}

Figure \ref{fig_b294skymap} shows a skymap with the location of the VVV b294 field (red filled square), to illustrate. 
According to the extinction map by \citet{Surot20}, this region presents a complex extinction structure. However, we do not expect nearby M dwarfs to be strongly affected, as most of the extinction is in the background. 
According to the 3D maps by \citet{Schultheis14}, the expected extinction is minimal for distances of less than 2\,kpc considering the line of sight of the b294 field. 
More recently, the 3D extinction maps from \citet{Lallement22} also show that for this region we have some minor extinction at about 200\,pc, however, it is mostly clear until about 1\,kpc, where denser clouds appear. 
In any case, extinction corrections are still considered in our analysis, as described later on section \ref{vosa}.

\section{Sample selection}\label{sample}

\subsection{Selecting candidates using parallax}\label{method1}

The objects from the b294 field -- with $\sim$4.6 million objects, as mentioned in the previous section -- were cross-matched using TOPCAT\footnote{TOPCAT is an interactive {\it Tool for OPerations on Catalogues And Tables}, available at \url{http://www.star.bris.ac.uk/~mbt/topcat/}} \citep{Taylor05,Taylor11} with the {\sl Gaia} Data Release 3 \citep[{\sl Gaia} DR3; ][epoch 2016]{Gaia2016,Gaia2022}, adopting a search radius of $1\arcsec$. Although high proper motion objects may lie outside the search region with such a small radius, it was chosen deliberately small to minimise the number of false counterparts, since we are dealing with a crowded field. From this search, we kept only around $\sim37$\% of the initial set ($1\,684\,116$ objects).

Only those objects with a good astrometric solution, adopted as  $\mathrm{RUWE} < 1.4$\footnote{The {\sl Gaia} Renormalised Unit Weight Error (RUWE) helps to identify non-single sources or objects with problematic astrometric solution. It is expected to be $\sim 1.0$ for single stars \citep{Arenou2018,Lindegren2018,Lindegren2021}. We selected objects with $\mathrm{RUWE} < 1.4$ as a compromise between sample completeness and minimum binary contamination.}, were selected, reducing the sample to $727\,342$ objects. 
We also kept only those that presented good parallax measurements ($\pi > 0$ and $\pi_{\rm err}/\pi < 0.2$; $42\,198$ objects). 
Keeping the relative errors below $20\%$ makes the inverse of the parallax a reliable distance estimator \citep{Luri18}. We also excluded those objects that showed {\sl Gaia} magnitude errors in $G$ and $G_\mathrm{RP}$ bands greater than $10\%$, resulting in a sample with $38\,483$ objects.

The final applied criterion was to keep only those stars within a distance of $500$ pc, in order to avoid high extinction levels expected for distant objects in the Galactic plane and, thus, effective temperature - extinction degenerancies in the Spectral Energy Distribution (SED) fitting. The obtained sample contains $2\,045$ M dwarf candidates. 

To identify the M-dwarf stars within this selection, we performed a SED fitting to obtain their effective temperatures (\teff). We used VOSA\footnote{\url{http://svo2.cab.inta-csic.es/theory/vosa/}} \citep[Virtual Observatory SED Analyzer,][]{Bayo2008} to obtain the stellar parameters based on broad-band photometry. The details of the fitting is further described in section \ref{vosa}. Those stars with estimated \teff\,$<4000$ K, and that presented a good fit solution (see Sect. \ref{vosa}), were kept as our first subset of candidates (hereafter, sample A) which contains $1\,338$ M-dwarf stars. 
Information on these objects can be found in the SVO archive of VVV M-dwarfs (see appendix \ref{app.cats}).

\subsection{Selecting candidates using proper motion and colour-cuts}\label{method2}

As an alternative way of selecting candidates, we applied several different filtering criteria to derive a sample of M-dwarf candidates.

Firstly, we kept only those objects with photometric data in all VISTA filters and with magnitude errors of less than $5\%$. This resulted in a subset of around 36\% of the initial sample ($1\,641\,542$ objects).

\citet{Rojas-Ayala2014} presented the expected colours for M dwarfs, based on VISTA filters. We applied their colour-cuts to our sample considering only the lower limits -- given for an M0 dwarf star -- as follows, which allowed us to reach later-type objects:

\begin{itemize}
\item[] $Y-J>0.336$;
\item[] $Y-H>0.952$;
\item[] $Y-K_{\rm s}>1.100$;
\item[] $J-H>0.432$;
\item[] $J-K_{\rm s}>0.642$;
\item[] $H-K_{\rm s}>0.045$.
\end{itemize}

\noindent
At this point, we identified $804\,452$ objects.

\subsubsection{Giant contaminants}

We performed a cross-match with {\sl Gaia} DR3 using TOPCAT, within a search radius of $1\arcsec$, to obtain the measured proper motions (hereafter, PM), $\mu$, for our objects. Here, we kept only those with PM errors on both directions, PM$_\mathrm{RA}$ and PM$_\mathrm{Dec}$, of less than $20\%$ (a total of $256\,890$ objects).  
Then, we calculated the reduced proper motion in VISTA J band, $H_{\rm J}$, using the definition from the seminal paper by \citet{Jones1972}\footnote{\citet{Jones1972} defined the reduced proper motion as $H = m+5\cdot\log \mu +5$, where $m$ is the apparent magnitude in a given photometric band.}. To discriminate between giant and dwarf star we adopted the following criterion described in \citet{Rojas-Ayala2014}:

\begin{equation}
H_{\rm J}^{d}> 68.5\cdot(J-K_{\rm s})-50.7,
	\label{eq_ReducedPM}
\end{equation}

\noindent
where $H_{\rm J}^{d}$ indicates the value of the reduced proper motion expected for a dwarf star as a function of $(J-K_{\rm s})$, colour that is calculated using VISTA magnitudes. 
Almost half of the objects in the sample ($130\,181$ objects) satisfied Eq.\,\ref{eq_ReducedPM}, which are those expected to be dwarf stars.

To minimise the amount of remaining giant contaminants, we applied another criterion based on the stellar proper motion. For that, we estimated the typical PM of M dwarfs using the SIMBAD Astronomical Database\footnote{ \url{http://simbad.u-strasbg.fr/simbad/}.} \citep{Wenger2000}, starting by the distribution of giant stars -- over $4\,200$ objects found, with a spectral type later than M0. We searched for these objects proper motions by cross-matching them with {\sl Gaia} DR3, keeping only those with good PM measurements (PM$_\mathrm{RA}$ and PM$_\mathrm{Dec}$ errors of less than $20\%$), resulting in $4\,043$. Analysing only those M giants (MIII) labelled as stars (SIMBAD OTYPE S $=$ 'Star') -- $2\,350$ of them --, we found an average PM $\sim 11.3$ mas/yr, where $\sim64$\% of the giant M-dwarf sample presented PM smaller than the average ($\sim60$\% if we consider $\mu$ $<10.0$\,mas/yr). 

Similarly, we searched for M dwarfs (MV) using SIMBAD, getting a sample of $20\,000$\footnote{This is in fact a subsample of M-dwarf stars from SIMBAD, since the tool limits the search to $20\,000$ results.} objects. Applying the same PM condition and selecting only the objects labelled as low-mass stars ($12\,646$; SIMBAD OTYPE S $=$ 'low-mass'), we derived an average $\mu$ $\sim 27.1$\,mas/yr. Adopting the PM average for MIII of $\sim 11.3$\,mas/yr, we found that $\sim76$\% of the tested dwarf sample has higher PM values ($\sim81$\% if we consider $\mu$ $>10.0$\,mas/yr). Hence, to maximise completeness of dwarfs and minimise contamination from giants, we set a cut in PM at $10.0$\,mas/yr.

To assess the performance of our whole approach for eliminating giants, we applied the related criteria -- good proper motion values ($err_{\rm \mu}/\mu<20\%$), cut in $H_{\rm J}$ from Eq. \ref{eq_ReducedPM}, and $\mu$ $\sim 10.0$\,mas/yr -- to the MIII sample from SIMBAD. From $4\,252$ giants (the complete MIII set), only $15$ giants survived the filtering, which represents $0.35\%$ of them. If considering only those giants labeled as stars ($2\,350$ objects), the remaining objects represent only $0.64\%$, meaning that these criteria were enough to eliminate $99.36\%$ of the giants. 
To verify the completeness of dwarf stars after applying these criteria, we repeated the giant filtering to the sample of MV from SIMBAD. We found that approximately half of the objects survived all criteria, which eliminated $50.6\%$ of the sample. 
Since we are prioritizing purity over completeness, the giant filtering is interpreted as a fair compromise between a good sample of dwarfs and minimum giant contamination.

Therefore, we kept only those objects that have their total proper motion higher than $10$ mas/yr. 
This way, we identified $24\,787$ candidate M-dwarf stars (or later) in our sample.

\subsubsection{Photometric distances}

From the list of $24\,787$ candidates, we kept those objects with good $G$ mag photometry (with relative error of less than 10\%) and with good astrometric solutions ($\mathrm{RUWE}<1.4$), resulting in a subset of $16\,652$ objects. 
We used the {\sl Gaia} and VISTA broad-band photometry to obtain the absolute $G$ magnitude, $M_{\rm G}$, as function of $(G-J)$ colour for our list of identified M dwarfs, considering the empirical relation described in \citet{Cifuentes2020}:

\begin{equation}
M_{\rm G} = 16.24-13.04\cdot(G-J)+5.64\cdot(G-J)^2-0.622\cdot(G-J)^3.
	\label{eq_linearFit}
\end{equation}

\noindent
This relation was derived using 2MASS J-band photometry. However, \citet{Gonzales2018} showed that the colour offsets in VISTA are modest and differences in magnitudes between 2MASS and VISTA are expected to be small, of around $5$ mmag in the J band\footnote{For an A0\,V star, the VISTA-2MASS colour in J band is $(J_{\rm VISTA}-J_{\rm 2MASS})=0.005\pm0.015$ \citep[][and references therein]{Gonzales2018}.}. It is also worth mentioning that the empirical relation (eq. \ref{eq_linearFit}) is valid for stars with certain values of $(G-J)$ colour, between $2.0$ and $4.0$. Applying this as condition, our sample is reduced to $16\,503$ objects.

The photometric distances were then computed from the estimated absolute G magnitudes and, as a final condition, only those stars with estimated distances of less than $500$ pc were selected -- $6\,846$ M-dwarf candidates.  
After performing a SED fitting with VOSA (for details, see Sect. \ref{vosa}), we kept only those which had presented a good fit solution and with \teff $< 4000$ K as we did in sect.\,\ref{method1}. This second subset (hereafter, sample B) contains $6\,747$ stars and it is also shown in the virtual observatory compliant archive described in Appendix \ref{app.cats}.

\subsection{The final sample}

We have compared the results obtained from the two methods previously described. Cross-matching both samples, we verified that only $160$ objects are common to both A and B samples. 
This is due to the different filtering conditions applied in each method, as explained in detail below.

\begin{figure}
    \centering
    \includegraphics[width=0.9\columnwidth]{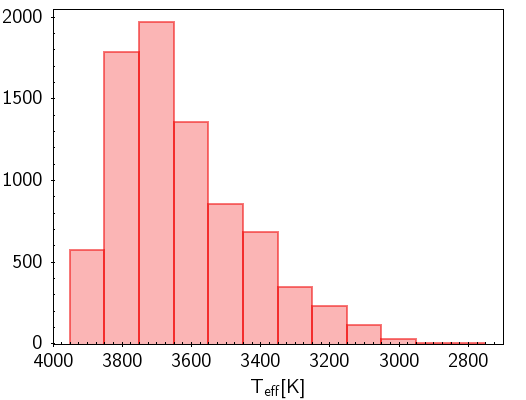}
    \caption{Distribution of effective temperatures obtained from the SED fitting with VOSA. The used grid of models has a step in temperature of 100\,K, which was adopted here as the bin size.}
    \label{fig_Teff}
\end{figure}

\begin{figure}
    \centering
    \includegraphics[width=0.9\columnwidth]{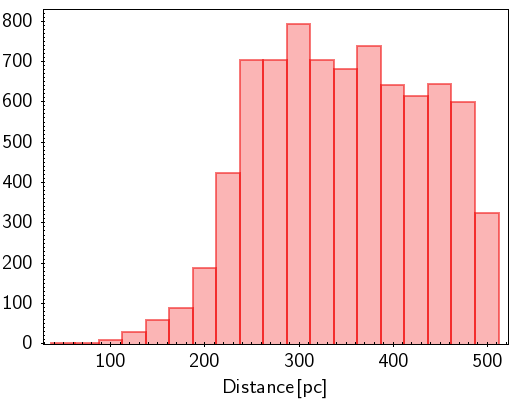}
    \caption{Distribution of obtained distances. The distances of M-dwarf candidates from sample A were obtained from {\sl Gaia} parallaxes. For sample B, photometric distances were adopted (see sect.~\ref{method1} and \ref{method2} for details).}
    \label{fig_dist}
\end{figure}

\begin{figure}
    \centering
    \includegraphics[width=0.97\columnwidth]{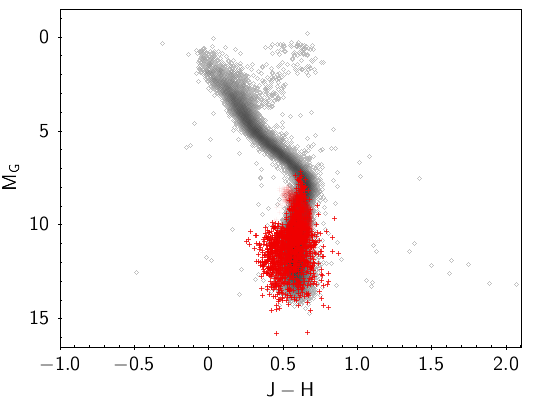}
    \caption{Colour-magnitude diagram showing the M dwarf sample from VVV b294 field. {\sl Gaia} nearby stars are represented by grey diamonds and red crosses show the M stars in our sample.}    \label{fig_CMD}
\end{figure}

By applying to sample A the same conditions considered for the second method -- which are colour-cuts for a M0 dwarf, photometric data in all VISTA/VVV bands with small error (less than $5\%$),  
cut in $H_{\rm J}$ (eq. \ref{eq_ReducedPM}), $\mu>10$\,mas/yr, and photometric distances of less than $500$ pc (see details in Sect. \ref{method2}) -- we recovered the same $160$ objects. Similarly, by applying to sample B the same criteria used in the first method -- which are good parallax measurements, photometric data in {\sl Gaia} DR3 $G$ and $G_\mathrm{RP}$ bands with small error, and distances from parallax of less than $500$ pc (see details in Sect. \ref{method1}) --, we found $160$ objects. 
Therefore, our final sample contains $7\,925$ ($1\,338$ + $6\,747$ - $160$) M-dwarf stars.  
Figures \ref{fig_Teff} and \ref{fig_dist} show the distribution of the estimated temperatures and distances of these objects. 

The identified M dwarfs are placed at the colour-magnitude diagram (CMD), as illustrated in figure~\ref{fig_CMD}. Grey diamonds represent a sample of {\sl Gaia} nearby stars from \citet{Torres2022}, after applying the quality metric described by \citet{Riello2021}\footnote{For details, see Table 2, and Eqs. 6 and 18 from \citet{Riello2021}.}, illustrating the main sequence locus. The colour $(J-H)$ for {\sl Gaia} stars was obtained from a cross-correlation with 2MASS, considering a $1\arcsec$ search radius. 
All objects from our sample of M stars are presented as red crosses.

\begin{figure*}
\centering
	\includegraphics[width=0.99\columnwidth]{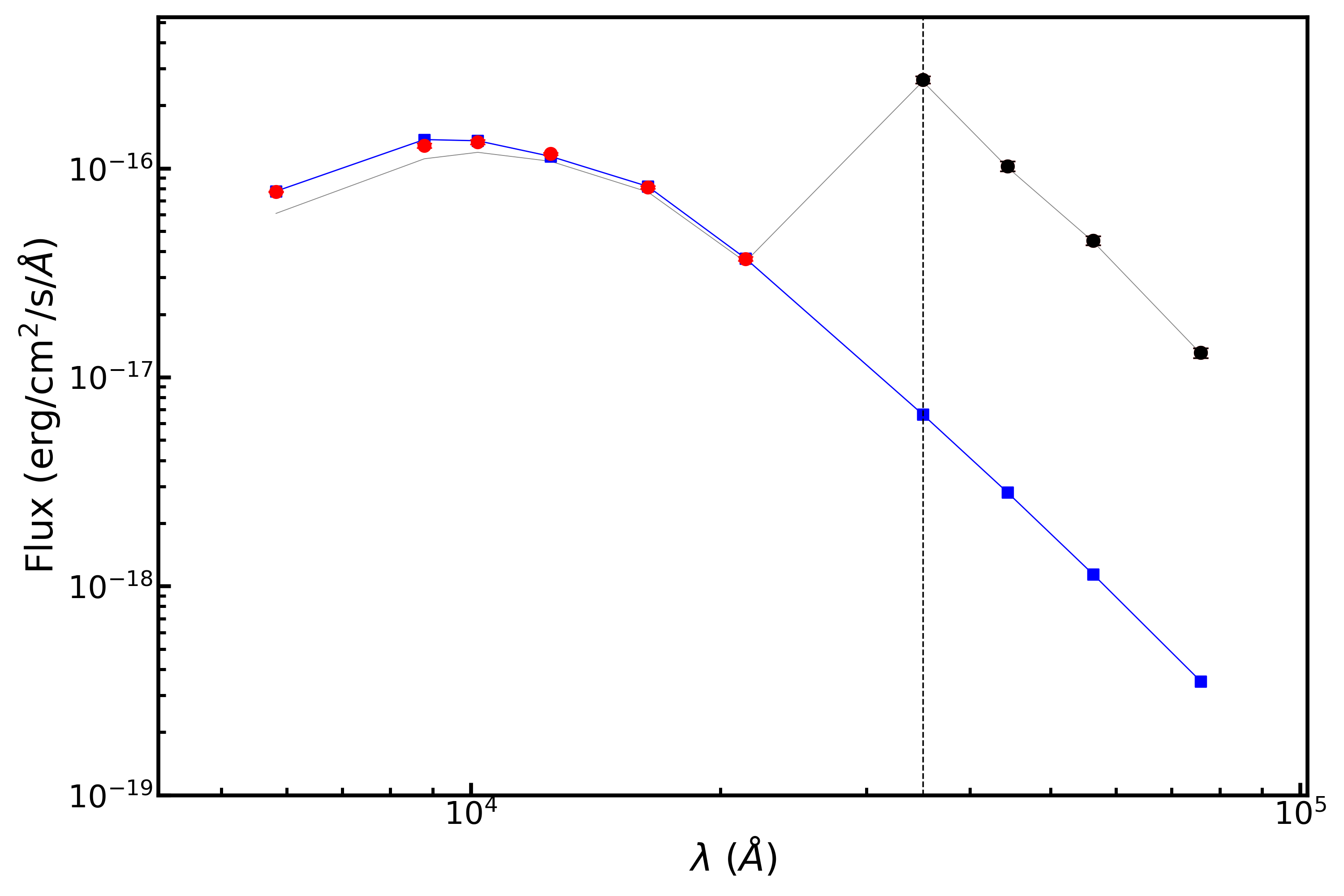}
	\includegraphics[width=0.99\columnwidth]{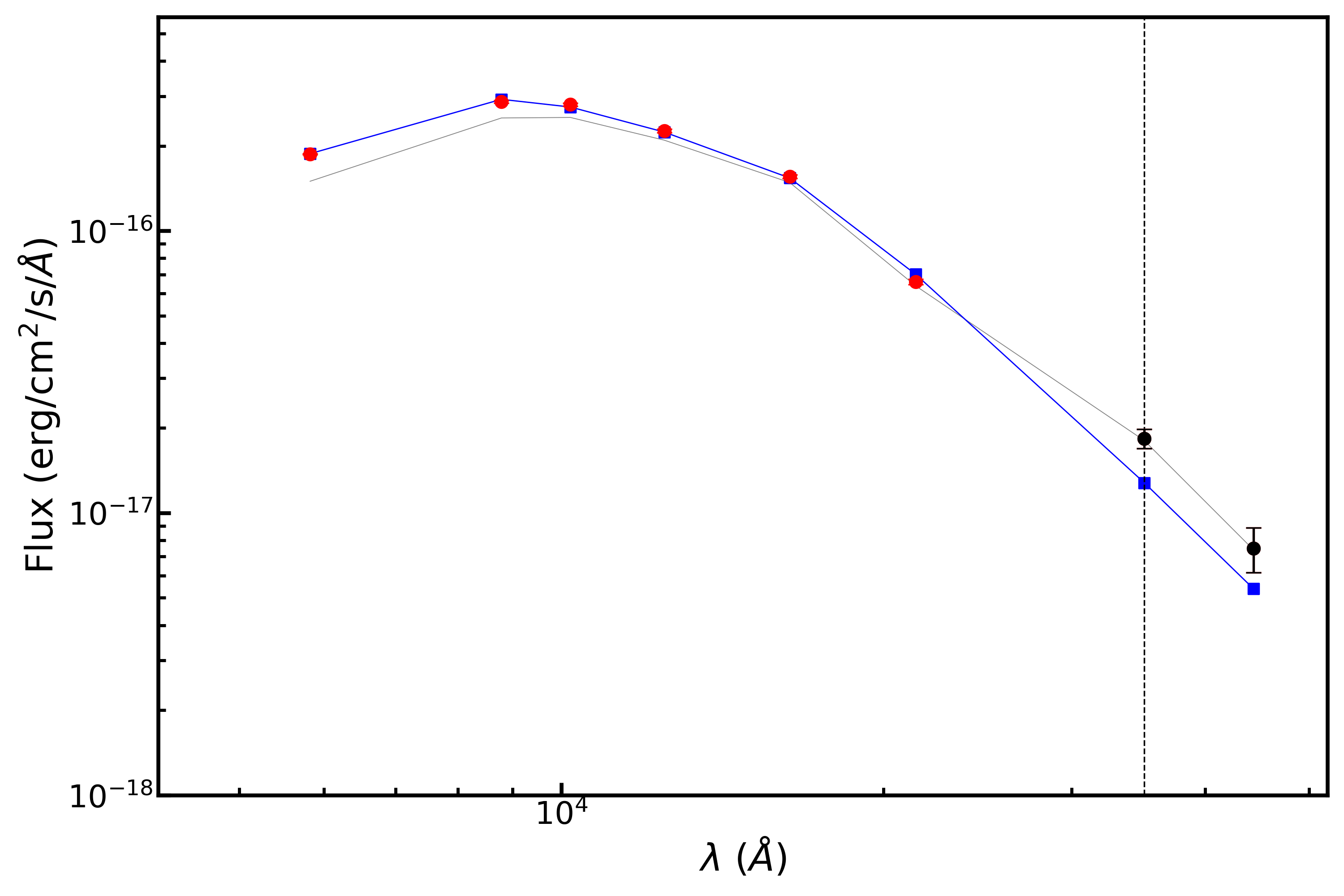}
 	\includegraphics[width=1.02\columnwidth,trim=1cm 1cm 1cm 1cm, clip]{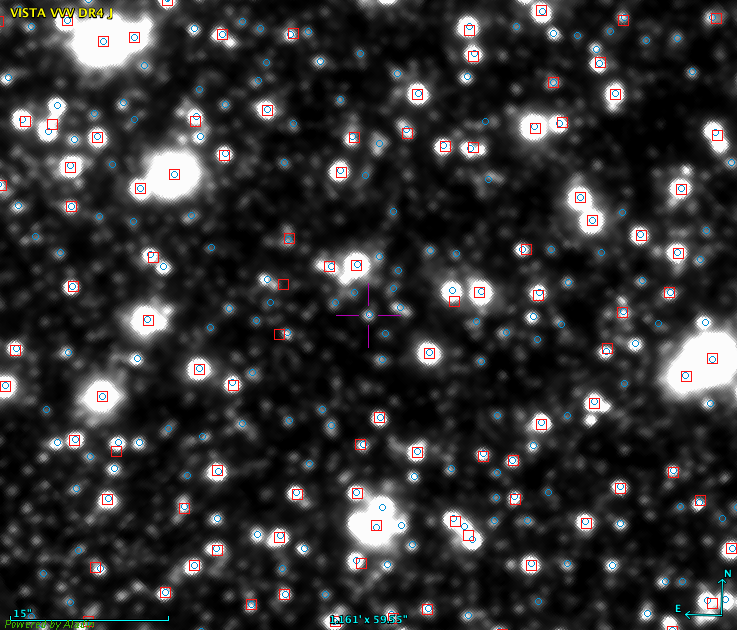}
	\includegraphics[width=1.02\columnwidth,trim=1cm 1cm 1cm 1cm, clip]{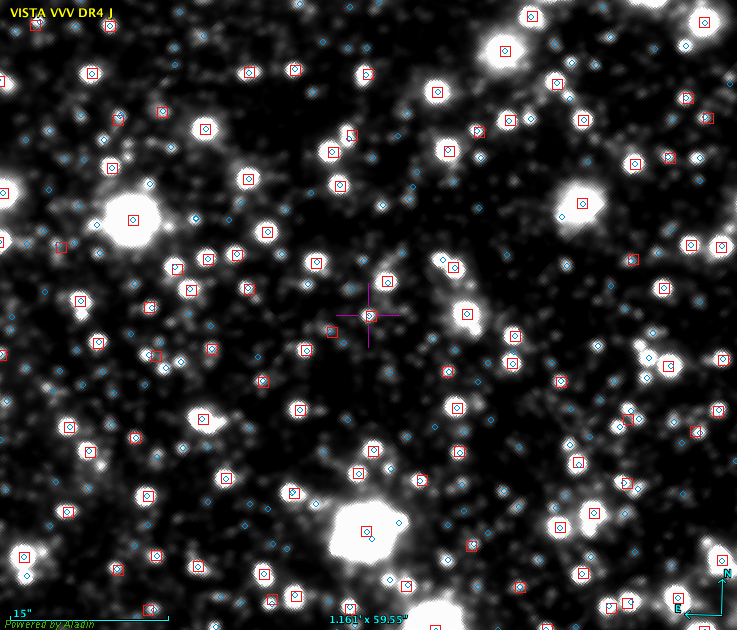}
    \caption{Example of two different cases of IR excess present from GLIMPSE photometry. The top panels show the results from SED fitting using VOSA and the identified starting point of IR excess (vertical dashed line), where blue squares represent the synthetic photometric calculated from the best-fit model, the red filled circles are the observed photometry, and black circles indicate the detected excess. The thin grey line shows the stellar SED before correcting from extinction. The lower panels (generated using Aladin) show the position of the same stars (marked as a pink cross at the center of each image) over a $J$-band image from VISTA/VVV survey. The objects identified in the VISTA/VVV catalogue are presented as small blue circles and those in the GLIMPSE catalogue are presented as red squares. The lower left panel illustrates an example of a star for which the GLIPMSE data came from a nearby object and, therefore, the IR excess in the SED (left upper panel) is not real. Different from the second example (right lower panel) where the detected excess is coming from the object.}
    \label{fig_excess}
\end{figure*}

\begin{table}
\caption{Broad-band photometry used for the SED fitting with VOSA. The information was compiled from the SVO Filter Profile Service \citep[][]{Rodrigo2012,Rodrigo2020}.}
\begin{tabular}{lccc}
\hline \hline
Band & $\lambda_{\rm eff}$ & $W_{\rm eff}$ & survey \\
name & (\AA) & (\AA) &  \\
\hline 
OmegaCAM u & 3607.68 & 482.54 & VPHAS+ DR2 \\
OmegaCAM g & 4679.46 & 1203.25 & VPHAS+ DR2 \\
{\sl Gaia} G & 5822.39 & 4052.97 & {\sl Gaia} DR3 \\
OmegaCAM H$\alpha$ & 6590.81 & 103.76 & VPHAS+ DR2 \\
OmegaCAM i & 7508.50 & 1463.68 & VPHAS+ DR2 \\
VISTA Z & 8789.53 & 889.46 & VVV \\
OmegaCAM z & 8884.40 & 864.48 & VPHAS+ DR2 \\
VISTA Y & 10196.43 & 870.63 & VVV \\
VISTA J & 12481.00 & 1542.53 & VVV \\
VISTA H & 16348.19 & 2674.02 & VVV \\
VISTA Ks & 21435.46 & 2793.85 & VVV \\
IRAC I1 & 35075.11 & 6836.16 & GLIMPSE \\
IRAC I2 & 44365.78 & 8649.93 & GLIMPSE \\
IRAC I3 & 56281.02 & 12561.17 & GLIMPSE \\
IRAC I4 & 75891.59 & 25288.50 & GLIMPSE \\
\hline \hline 
\end{tabular}
\label{tab_broadband}
\end{table}

\section{Stellar properties with VOSA}\label{vosa}

As briefly mentioned in Sect. \ref{sample}, we used VOSA \citep{Bayo2008} to obtain effective temperatures, luminosities and radii for all M-dwarf candidates in our selected sample. VOSA is a tool developed by the Spanish Virtual Observatory\footnote{\url{http://svo.cab.inta-csic.es}} designed to build the SEDs of thousands of objects at a time from a large number of photometric catalogues, ranging from the ultraviolet to the infrared. VOSA compares catalogue photometry with different collections of theoretical models and determines which model best reproduces the observed data, following different statistical approaches. Physical parameters are then estimated for each object from the model that best fits the data.

To construct the SED, we used the VISTA $ZYJHK_s$ photometry obtained from the VVV survey. We also searched for additional broad-band photometry from {\sl Gaia} DR3 (only in the $G$ band), from the VST Photometric H${\rm \alpha}$ Survey of the Southern Galactic Plane and Bulge \citep[VPHAS+ DR2;][]{Drew2014}, and from the Galactic Legacy Infrared Mid-Plane Survey Extraordinaire \citep[GLIMPSE Source Catalog I + II + 3D;][]{Spitzer2009}. Information on the filters is shown in table \ref{tab_broadband}.

It is worth mentioning that only a small portion of the sample (172 objects) had VPHAS+ DR2 magnitudes available. For these objects, the colour $(r-H\alpha)$ from VPHAS+ DR2 was used to search for H$\alpha$ emission. We found that all the objects with $r$ and H$\alpha$ photometry available have $(r-H\alpha)$ < 0.9, which is expected for non-emitting M sources \citep{Drew2014}.

The BT-Settl CIFIST models -- the BT-Settl theoretical spectra by \citet{Allard2011} computed with a cloud model and using the \citet{Caffau2011} solar abundances -- were adopted for the SED fitting, covering effective temperatures from $1200$ to $7000$ K and assuming solar metallicity (${\rm [Fe/H]=0.0}$). Since we expect our candidates to be dwarf stars (see Sect. \ref{sample} for details), the surface gravities were allowed to vary only from $4.0$ to $5.5$. Finally, we defined a range of possible values for the extinction ($A_{\rm v}$), going from $A_{\rm v}=0.0$ to $0.5$, as expected according to extinction models for the region. 
Extinction is left as a free parameter to be fitted together with the stellar parameters in the SED fitting process. To calculate the extinction in each filter, VOSA uses the extinction law by \citet[][]{Fitzpatrick1999}, with the improvement in the infrared region by \citet[][]{Indebetouw2005}\footnote{For more details on the interstellar extinction, see VOSA's help section at \url{http://svo2.cab.inta-csic.es/theory/vosa/help/star/extinctions/}.}.

After the fit is performed, we opted to refine the results due to excess present in the infrared data.  
By comparing the photometric data to the best-fit model, VOSA is able to identify which data points appear high above the model and (re)define the starting point of the infrared excess. We then refitted those SEDs that VOSA identified to have an excess, where all the points which were found to be affected were not considered in the final fit.

Since the b294 tile is placed in a crowded field, some of the detected IR excess could just be due to a wrong counterpart assignment. This is illustrated in fig. \ref{fig_excess}. On the left, the example of an object with a SED showing a strong IR excess (top left panel) but later verified to come from a different source (bottom left panel). For comparison, a more subtle excess in the SED of a different object is presented on the right (top right panel), later confirmed to come from the same emitting source (bottom right panel). 
All objects for which VOSA has identified an infrared excess are properly flagged in the archive, as described in Appendix~\ref{app.cats}. These detected IR excess need to be further confirmed, as explained above.

We obtained a good fit for all objects in our sample, according to visual goodness-of-fit value ($V{\rm gf_b}$) estimated by VOSA, which is the modified reduced $\chi^2$ calculated by forcing the error in the observed flux to be larger than $10$\%\footnote{The visual goodness-of-fit value, $V{\rm gf_b}$, smaller than 12--15 is often perceived as a good fit. For more details, see VOSA's help page \url{http://svo2.cab.inta-csic.es/theory/vosa/help/star/fit/}.}. 
The results of the SED fitting are also included in the archive described in Appendix \ref{app.cats}.

\section{Comparison to {\sl Gaia} M dwarfs}\label{discuss}

\begin{figure}
    \centering
    \includegraphics[width=0.98\columnwidth]{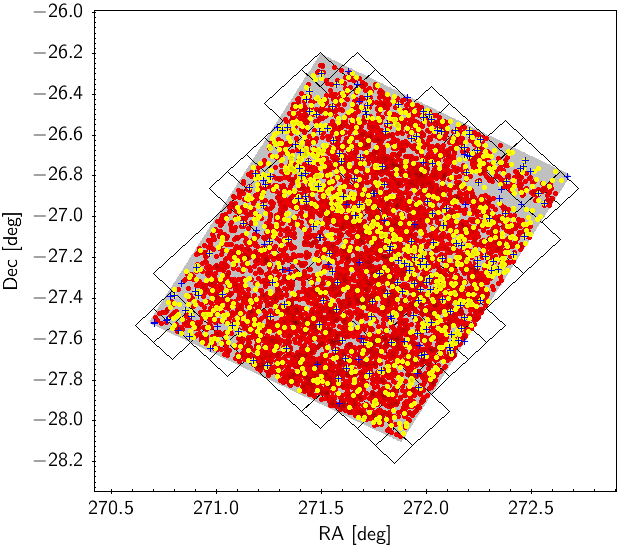}
    \caption{Sky position of the $7\,925$ M-dwarf stars in our sample. The VVV tile b294 is presented as a grey area and the solid black lines represent its MOC. Blue crosses show the M dwarfs identified in {\sl Gaia} DR3 within 500\,pc.}
    \label{fig_Mdwarf_Gaia}
\end{figure}

\begin{figure}
    \centering
    \includegraphics[width=1.\columnwidth]{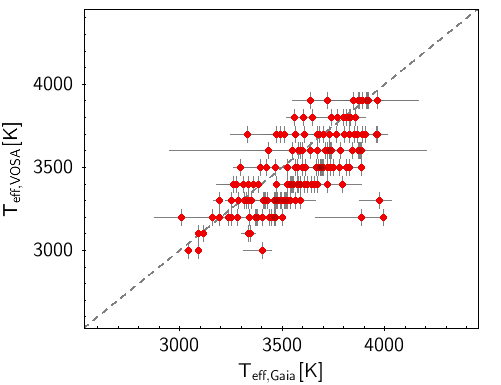}
    \caption{Comparison of \teff obtained in our analysis (VOSA) with those given in {\sl Gaia} DR3, for the stars identified as M dwarfs in both sets. The dashed line represents the identity function.}
    \label{fig_Teff_Gaia}
\end{figure}

The sky distribution of the $7\,925$ M-dwarf stars identified in VVV tile b294 is presented in figure\,\ref{fig_Mdwarf_Gaia}, where the grey area represents the observed field of view and the black solid line shows the Multi-Object Coverage (MOC). The objects from samples A and B are represented by yellow and red filled circles, respectively.

To assess the importance of our methodology and the impact of our M stars catalogue to the identified cool stars in the studied region -- towards the Galactic bulge (see fig. \ref{fig_b294skymap}) -- we exploited the available data from {\sl Gaia} DR3 to search for known M dwarfs in the VVV tile b294 field. From the approximately 2 million sources of the {\sl Gaia} DR3 present in tile b294 field (fig. \ref{fig_Mdwarf_Gaia}, grey area), we selected those with RUWE of less than 1.4 -- the same limit applied for VVV objects and described in sect.~\ref{sample} -- to ensure a good astrometric solution for a point-like source. As a matter of consistency, we selected the stars with a relative parallax error of less than 20\%, keeping those within a distance of 500\,pc. Aiming at identifying M dwarfs, we kept those stars with \teff\ of less than $4000$\,K and $\log{g}\ge4.0$ -- astrophysical parameters available in {\sl Gaia} DR3 catalogue, which were derived from BP/RP spectra. The resulting sample contained only $208$ stars characterised in {\sl Gaia} DR3 as M-dwarf stars. All these {\sl Gaia} DR3 M stars are shown in fig.\,\ref{fig_Mdwarf_Gaia} as blue crosses. 
Therefore, we have increased considerably the number of M dwarfs identified in the field, from a few hundreds to thousands of stars. 

Figure~\ref{fig_Teff_Gaia} shows the comparison between the effective temperatures derived from SED fitting with VOSA and the ones given in {\sl Gaia} DR3 catalogue for the $162$ objects in common. Considering the estimated uncertainties, there is in general a good agreement between derived values, with a mean difference of $181$\,K and a standard deviation of $136$\,K, approximately. Among the M stars identified in {\sl Gaia} DR3, $39$ of them were not included in the cross-match with the original VVV tile b294 catalogue, considering $1\arcsec$ radius, and $7$ were lost in the applied criteria, such as colour-cuts and proper motion limits. 
 
Our sample has increased significantly the number of known M dwarfs in the studied region -- a crowded field towards the bulge of the Galaxy -- emphasising the importance of ground-based photometric surveys in the near-infrared.

\section{Searching for periodic signals in VVV light curves}\label{period}

As a secondary outcome of this study, it is interesting to analyse the multi-epoch observations performed in VVV tile b294 and search for periodic signals.

The time series were obtained with the VVV $K_s$ filter, with over $300$ observations of our field of interest. Not only this, but due to the strategy of observation \citep{Saito12}, every region of the field is observed at least twice, most of the times using different chips of the 16-chip VIRCAM instrument used in the VISTA telescope. For our multi-epoch analysis described below, we made use of VVV $K_s$-band light curves based on PSF photometry as described in \citet[][]{Contreras17}.

\subsection{Light curve selection}\label{sec_lcselection}

We gathered light curves with more than 25 detections for 7\,752 M dwarfs out of the $7\,925$ stars in our sample. Some of them have been detected in more than one chip due to overlap, providing a total of 8\,640 different light curves in the $K_s$/VISTA band. The time-series data covered in total 5.4\,yr of observations, acquired between 12 April 2010 and 11 September 2015 \citep[][]{Contreras17}.

In order to clean spurious data present in the light curves, we carried out a two-step procedure:
\begin{enumerate}
    \item We ran a sigma-clipping approach aiming at identifying the baseline emission of the star.
    \item From the original light curve, we kept all epochs within the first and 99th centiles and discarded those brighter than the median value minus 3$\sigma$, computed from the set obtained in the previous exercise. These detections could be ascribed as mismatches that could contaminate the search for periodic signals. 
\end{enumerate}
In order to keep the sample purity, outliers were excluded from the light-curve analysis.

\begin{figure}
    \includegraphics[width=0.9\columnwidth,trim=0cm 0cm 0.5cm 0cm, clip]{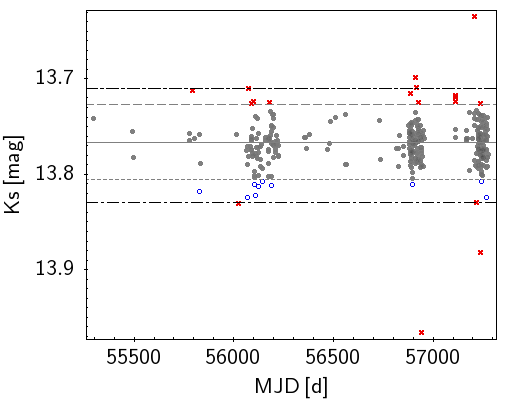}
    \includegraphics[width=0.9\columnwidth,trim=0.1cm 0cm 0.4cm 0cm, clip]{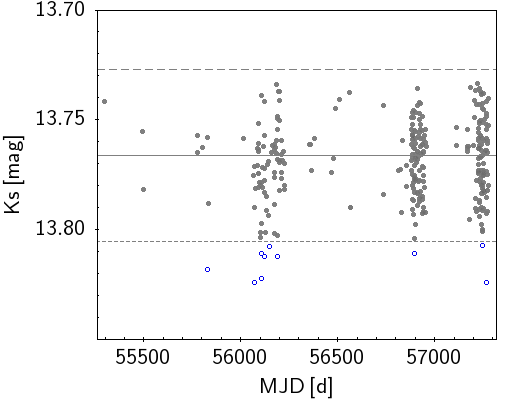}
    \caption{{\it Top panel:} Example light curve. Baseline emission data points, detections related to variability and rejected outliers, which could also be related to flares, are displayed in grey filled circles, blue open circles and red crosses, respectively. Grey solid, short dashed and long dashed lines represent the median, the median+3$\sigma$ and median-3$\sigma$ limits, respectively. Black dashed-dotted lines represent the first and 99th quartiles. {\it Bottom panel:} Light curve of the same object after removing outliers. Colour and symbol code as in the figure on top.}
    \label{fig_lc}
\end{figure}

All light curves that present more than three points with magnitude values between the median plus 3$\sigma$ and the 99th percentile were considered suitable for the analysis. A drop in brightness can be ascribed to either a transiting object or to starspots present in the stellar surface of active M dwarfs. 
We obtained 8\,219 light curves associated to 7\,153 M dwarfs. An example of a raw light curve is shown on top panel in Figure~\ref{fig_lc}.

Additionally, we applied a photometric quality filter and discarded all the light curves with mean magnitude errors larger than 1.5 times the Median Absolute Deviation (MAD) obtained from the baseline emission of each star in step (i). After this, 4\,952 stars with 5\,817 light curves remained. They were normalised to the median and combined in order to obtain a unique light curve per star.

\subsection{Light curve analysis}\label{LCanalysis}

Although the light curves span over 5.4\,yr, there are three main observing blocks of nearly 6.5 months each (200\,d), going from May to November 2012, from May to December 2014, and from March to October 2015. 
With the goal of providing a reliable set of M-dwarf candidates with trustful periodic variations, we searched for periodic signals in each of the three main blocks of observations rather than in the whole data set. Figure~\ref{fig_lc} shows an example of a whole light curve where the three observing blocks used for deriving the periods are clearly differentiated.

Given the unevenly spaced observations, we obtained the periodograms from each block in the light curves using two complementary methods. Firstly, we applied the Generalized Lomb-Scargle algorithm \citep[GLS,][]{Lomb76,Scargle82}, which fits a sinusoidal model to the data at each frequency. For that, we employed the Python {\tt astropy.timeseries} package \citep{astropy:2013,astropy:2018} and obtained the best periods (i.e., the periods with the highest peak in the periodogram). We tested them through the false alarm probability \citep[FAP,][]{Scargle82} computed with the Baluev approximation \citep{Baluev08}. We derived three periods -- one period per block -- but only periods with FAP under 0.1 were taken into consideration. 
Secondly, we searched for periodic signals applying the Phase Dispersion Minimization method \citep[PDM,][]{PDM78}, between 0.1 and 100\,d. Unlike GLS, PDM is unbiased towards the shape of the light curve and is able to find also non-sinusoidal variations.

\begin{figure}
\centering
    \includegraphics[width=0.85\columnwidth,trim=0cm 0cm 0.5cm 0cm, clip]{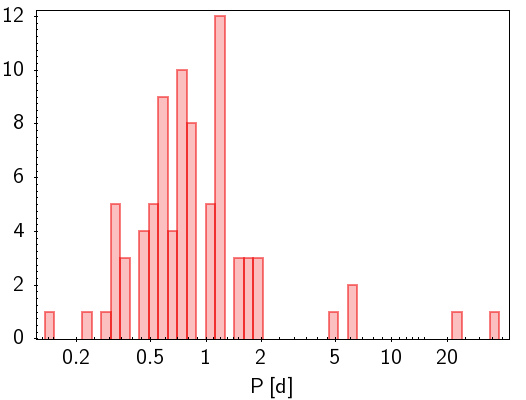}
    \caption{Period distribution in logarithmic scale for the sample of 82 M dwarfs.}
    \label{fig_Pdist}
\end{figure}

We obtained 224 periods in agreement within 20\% from the LSG and PDM approximations. After removing near 1\,d periodic signals (from 0.9 to 1.1\,d), 82 stars remain. Additionally, we use the reduced chi-square as a variability indicator as in \cite{Botan21} and define two subsets of M dwarfs with periodic signals. One is composed by 27 M dwarfs with $\chi^2 > 2$, whose variability would likely be related to the presence of a stellar companion, indicating possible binary systems. The obtained periods range between 0.14 and 34\,d. Among them, $2$ objects were recently identified as variables in a recent work by \citet{Molnar22}, where they were flagged as eclipsing binary systems. 
The other subset is composed by 20 M dwarfs that present $\chi^2 \sim 1$ (from 0.7 to 1.3), for which the periodic signal could be ascribed to a planet-like transiting object. However, the found periodicities could also be related to stellar intrinsic variability, due to starspots for instance. Therefore, the periods need to be confirmed (or discarded) with dedicated follow-up observations. The periods distribution is shown in Figure~\ref{fig_Pdist}. Each of the objects for which periodicity has been calculated is properly flagged in the archive, as described in Appendix~\ref{app.cats}.

\section{Conclusions}\label{concl}

We searched for M-dwarf stars near the Galactic bulge in the b294 field from the VISTA Variables in the Vía Láctea survey. We adopted two different methodologies to identify M dwarfs. The first method was performed using parallaxes, where we selected objects with good astrometric solution, photometry -- relative G and RP magnitude errors below 10\% -- and parallax measurements -- relative error of less than 20\% -- from {\sl Gaia} DR3. 
The second method was based on colour-cuts and proper motions, where we kept those objects with good VISTA photometry in all $YZJHK_s$ bands, VISTA colours among the colour-cuts defined by \citet{Rojas-Ayala2014} for M stars, good proper motion -- relative error of less than 20\% -- and good astrometric solution also from {\sl Gaia} DR3, and those with a J magnitude reduced proper motion expected for dwarf stars (see Sect.~\ref{method2} for details). 
We then estimated absolute magnitudes from the empirical relation based on colours by \citet{Cifuentes2020}.

To avoid the interstellar extinction expected for the field, we selected only the M-dwarf candidates within 500\,pc, where distances where estimated from parallax (sample A) and photometric distances based on absolute magnitudes (sample B). 
We then characterised the remaining candidates by performing SED fittings using VOSA, where we kept all objects with \teff\,$<$ $4\,000$ K. Our final list of M-dwarf candidates from the VVV b294 field has $7\,925$ stars, with temperatures ranging from $2\,800$ to $3\,900$ K.

To assess the importance and impact of the identified M stars towards the Galactic bulge, we compared our sample to all M dwarfs characterised from BP/RP spectra available in {\sl Gaia} DR3 catalogue in the VVV tile b294 field. From nearly 2 million sources, there are $208$ stars with \teff $ $ and $\log{g}$ compatible with M-dwarf stars. 
Our sample of $7\,925$ sources has significantly increased the number of known M dwarfs within 500\,pc in the studied region. 

As a secondary outcome of this study, we also searched for periodic signals in VVV light curves, with at least 25 and up to 327 epochs. We removed outliers from the light curves and looked for periodicities in the three main blocks of observation. We obtained periods for 82 M dwarfs by applying two methods: the Lomb-Scargle and Phase Dispersion Minimization, independently. These periods range from 0.14 to 34\,d. We defined two subsamples according to the reduced chi-square (see Sect.~\ref{LCanalysis}) presenting large and small variability (27 and 20 stars, respectively). Additional follow-up observations and further analysis would be required for confirming the nature of the periodic variability of these objects. 

Even with the amazing collection of data delivered by {\sl Gaia} DR3, radial velocities and spectra are not available for all observed objects. The methodology described in this work probed to be very efficient on identifying and characterising M-dwarf stars in the VVV b294 field, emphasising the importance of ground-based photometric surveys in the near-infrared. 
Therefore, it can be extended to other VVV fields -- and to those from the VVVX survey \citep{Minniti18} -- in order to increase the population of known low-mass objects in the direction of the Galactic bulge.

\section*{Acknowledgements}

We would like to thank Dr F. Jiménez-Esteban for the fruitful discussion. 
P.C. acknowledges financial support from the Government of Comunidad Autónoma de Madrid (Spain) via postdoctoral grant `Atracción de Talento Investigador' 2019-T2/TIC-14760. 
M.C.C. acknowledges financial support from the ESCAPE project supported by the European Commission Framework Programme Horizon 2020 Research and Innovation action under grant agreement n. 824064.
This research has made use of the Spanish Virtual Observatory (\url{https://svo.cab.inta-csic.es}) project funded by the Spanish Ministry of Science and Innovation/State Agency of Research MCIN/AEI/10.13039/501100011033 through grant PID2020-112949GB-I00 and MDM-2017-0737 at Centro de Astrobiología (CSIC-INTA), Unidad de Excelencia María de Maeztu. 
D.M. also thanks the support by the ANID BASAL projects ACE210002 and FB210003, and Fondecyt Project No. 1220724. 
J.A.-G. acknowledges support from Fondecyt Regular 1201490, and ANID – Millennium Science Initiative Program – ICN12\_009 awarded to the Millennium Institute of Astrophysics MAS. 
R.K.S. acknowledges support from CNPq/Brazil through project 305902/2019-9. 
We gratefully acknowledge the use of data from the ESO Public Survey program IDs 179.B-2002 and 198.B-2004 taken with the VISTA telescope and data products from the Cambridge Astronomical Survey Unit. 
We would also like to thank R. Contreras Ramos (private communication) for the VVV light curves used in the present work.

This publication makes use of VOSA, developed under the Spanish Virtual Observatory project. VOSA has been partially updated by using funding from the European Union's Horizon 2020 Research and Innovation Programme, under Grant Agreement nº 776403 (EXOPLANETS-A).
This research has made use of the SVO Filter Profile Service (\url{http://svo2.cab.inta-csic.es/theory/fps/}). This research has made use of "Aladin sky atlas" developed at CDS, Strasbourg Observatory, France. This research has made use of the SIMBAD database, operated at CDS, Strasbourg, France.

\section*{Data Availability: Virtual Observatory compliant, online catalogue}\label{archive}

In order to help the astronomical community on using our catalogue of VVV M dwarfs, we developed an archive system that can be accessed from a webpage\footnote{\url{http://svocats.cab.inta-csic.es/mdwarfs_vvv/}} or through a Virtual Observatory ConeSearch\footnote{A ConeSearch example can be seen at \url{http://svocats.cab.inta-csic.es/mdwarfs_vvv/cs.php?RA=271.877&DEC=-28.079&SR=0.1&VERB=2}}. The content of the catalogue is presented in the Appendix~\ref{app.cats}.

The archive system implements a very simple search interface that allows queries by coordinates and radius as well as by other parameters of interest. The user can also select the maximum number of sources (with values from 10 to unlimited) and the number of columns to return (minimum, default, or maximum verbosity). 
The result of the query is a HTML table with all the sources found in the archive fulfilling the search criteria. The result can also be downloaded as a VOTable or a CSV file. Detailed information on the output  fields can be obtained placing the mouse over the question mark located close to the name of the column. The archive also implements the SAMP\footnote{\url{http://www.ivoa.net/documents/SAMP}} (Simple Application Messaging) Virtual Observatory protocol. SAMP allows Virtual Observatory applications to communicate with each other in a seamless and transparent manner for the user. This way, the results of a query can be easily transferred to other VO applications, such as, for instance, TOPCAT.




\bibliographystyle{mnras.bst} 
\bibliography{Mdwarfs_VVV.bib} 




\appendix

\section{Catalogue description}\label{app.cats}

The content of the catalogue of M-dwarf candidates from the VVV b294 field is presented in Table~\ref{tab.catalogue_description}. 
This catalogue can be accessed from the dedicated webpage \url{http://svocats.cab.inta-csic.es/mdwarfs_vvv/} or through a Virtual Observatory ConeSearch (e.g. \url{http://svocats.cab.inta-csic.es/mdwarfs_vvv/cs.php?RA=271.877&DEC=-28.079&SR=0.1&VERB=2}).

\begin{table*}
\centering
\caption{Description of the parameters contained in the VVV M-dwarf catalogue.}
\label{tab.catalogue_description}
\begin{tabular}{lcl}
\hline \hline
	\noalign{\smallskip}
Parameter    &   Units   &   Description \\
	\noalign{\smallskip}

	\hline
Gaia\_ID\_DR3	&	-	&	{\sl Gaia} DR3 source identifier.	\\
RAJ2000	&	deg	&	Celestial Right Ascension (J2000).	\\
DEJ2000	&	deg	&	Celestial Declination (J2000).	\\
Xmag    &   mag & Calibrated magnitude. {\it X} stands for $Z$, $Y$, $J$, $H$ and $K_s$.    \\
eXmag   &   mag & Calibrated magnitude error. {\it X} stands for $Z$, $Y$, $J$, $H$ and $K_s$.  \\
d	&	pc	&	Distance.	\\
Ref\_d   &   -   &   Reference for the distance. "Gaia" refers to parallactic distances and "This work" indicates distances are \\
&&specto-photometric derived in this work.\\
\teff   &   K   & Effective temperature. \\
L$_{\rm bol}$  & L$_\odot$  &      Bolometric luminosity. \\
eL$_{\rm bol}$  & L$_\odot$  &      Error in the bolometric luminosity. \\
R   &    R$_\odot$  & Stellar radius. Calculated using L$_{\rm bol} = 4 \pi R^2 \sigma$ \teff$^4$. \\
eR   &    R$_\odot$  & Error in the stellar radius. \\
Av  & -  &  Visual extinction. \\ 
Method  &   -   & Identification method of candidates. "A" stands for parallax, and "B" stands for proper motion and colour-cuts. \\
Flag\_IR &   -   & Flag for stars with IR excess detected by VOSA yet to be confirmed.    \\ 
Flag\_lc &   -   & Flag for dwarfs with processed light curve.    \\ 
Flag\_P &   -   & Flag for dwarfs with periodic signal.    \\ 
Flag\_chi  &   -   & Flag for M dwarfs with $\chi^2 \sim 1$ (1) and $\chi^2 > 2$ (2). \\

	\noalign{\smallskip}

\hline
\end{tabular}
\end{table*}



\bsp	
\label{lastpage}
\end{document}